\documentclass{IEEEtran}

\usepackage{graphicx}
\usepackage{amsfonts}
\usepackage{multicol}
\usepackage{float}
\usepackage{algorithm}
\usepackage{algorithmic}
\usepackage{amsmath}
\usepackage{amsthm}
\usepackage{booktabs} 
\usepackage{cite}
\usepackage{color}

\graphicspath{{./Figs/}}
\definecolor{BoxBackground}{rgb}{0.74, 0.83, 0.9}
\DeclareMathOperator*{\argmin}{\arg\!\min} 
\newcommand{\argmax}{\arg\!\max}
\renewcommand{\v}[1]{{\bf{#1}}}

\begin{document}

\title{Compressed Ultrasound Imaging:\\
from Sub-Nyquist Rates to Super-Resolution}

\author{Oded Drori, Alon Mamistvalov, Oren Solomon, Yonina C. Eldar,~\IEEEmembership{Fellow,~IEEE}}

\maketitle


\section{Introduction}
\label{sec:intro}
The multi-billion dollar, worldwide medical ultrasound (US) market continues to grow annually. Its non-ionizing nature, real-time capabilities and relatively low cost, compared to other imaging modalities, have led to significant applications in many different fields, including cardiology, angiology, obstetrics and emergency medicine. Facilitated by ongoing innovations, US continues to change rules and norms regarding patient screening, diagnosis and surgery. This huge and promising market is constantly driven by new imaging and processing techniques. From 3D images to sophisticated software, hardware and portability improvements, it is clear that the status of US as one of the leading medical imaging technologies is ensured for many years ahead. However, as imaging systems evolve, new engineering challenges emerge. Acquisition, transmission and processing of huge amounts of data are common for all ultrasound-based imaging modalities. Moreover, achieving higher resolution is constantly on demand, as improved diagnosis could be achieved by better visualization of organs and blood vessels deep within tissues.

In this article, our goal is to motivate further interest and research in emerging processing techniques, as well as their applications in medical ultrasound, enabled by recent advancements in signal processing algorithms and deep learning \cite{lecun2015deep}. 
We address some of the primary challenges and potential remedies from a signal processing perspective, by exploiting the inherent structure of the received US signal ~\cite{eldar2012compressed, eldar2015sampling}. In particular, we discuss the mathematical models which underlie the received signals and show how prior information can be exploited to achieve unprecedented improvements in both data acquisition trade-offs and resolution. 
We begin by detailing the processing steps necessary to obtain an ultrasound image in the most commonly used schemes today. 
Once these are established, we turn to the challenges inherent to this process, focusing on data rates, access to the actual signal returns, and resolution.

The first challenge we address is that of the data-rate bottleneck in US imaging. The formation of a US image requires large amounts of data due mostly to two factors: High sampling rates, multiples or even magnitudes of order beyond the Nyquist rate to ensure high resolution beamforming, and a large number of data channels (and corresponding receivers), typically tens to hundreds, to enable high spatial resolution. These high data rates constrain advanced processing algorithms and portable hardware \cite{eldar2015sampling}. Typically, to enable real time imaging, a significant amount of the processing is done in hardware, with only reduced-dimensional digital images being exported from the device. This inability to access the original, raw signals (`channel data') limits the ability to take advantage of advanced, digital processing and machine learning methods \cite{van2019deep}. Additionally, processing in hardware means that bulky and costly systems must be used, housing all required electronics, limiting portability and increasing cost. Currently, whenever reducing system costs becomes a priority, an unnecessary reduction in performance follows.

To address the problem of data-rate reduction, we review methods brought upon by the ideas of compressed sensing ~\cite{eldar2012compressed,tur2011innovation,eldar2015sampling}, where prior knowledge about the frequency content and structure of ultrasound signals may be used to represent it using a reduced set of samples. In particular, we exploit the fact that the recived signal is sparse, centered around the transmitted frequency, and that the beamformed signal used to form an image corresponds to time-shifted and
scaled reflections of the transmitted pulse, thus obeying the Finite Rate of Innovation model \cite{chernyakova2013compressed}. The result is an ability to reconstruct an equivalent image using fewer samples, reducing required sampling rates by orders of magnitude, even below the Nyquist limit of the transmitted pulse \cite{chernyakova2013compressed}.

To reduce the number of channels used in US probes, we turn to the field of sparse array design ~\cite{werner2003overview,cohen2020sparse}. 
With an appropriate choice of sparse array and post processing, the performance of a large physical array, as characterized by the beam pattern, can be achieved using fewer elements enabling a reduction in the number of elements without loss of performance. 
Algorithms such as Convolutional Beamforming \cite{cohen2018sparse}, can be used to reconstruct an equivalent image with four to ten times fewer channels. 

Addressing the data bottleneck leads to the availability of US channel data, enabling the application of emerging deep learning techniques directly on the raw channel returns. This approach can improve a seemingly limitless array of applications, including noise reduction, enhanced resolution, advanced diagnostic tools, addressing technician dependence,  and rapid image formation, to name a few ~\cite{yoon2018efficient,van2019deep}. We demonstrate two such deep learning algorithms: the LISTA algorithm \cite{gregor2010learning}, used to perform efficient recovery of an ultrasound signal from a small set of samples given its sparse representation \cite{mamistvalov2021deep}, and the ABLE algorithm \cite{luijten2020adaptive}, used to form high resolution US images from their channel data. Moreover, reducing the data size and computational load needed in today's traditional US machines, paves the way for over-WIFI, efficient, wireless US imaging, bringing US machines closer to the patients, independent of their geographical location.

Finally, we turn to address the challenge of increasing US image resolution \cite{ackermann2016detection}. As a wave-based imaging modality, conventional spatial resolution for US signals is on  the order of its wavelength, ranging from a few tenths of a millimeter to a few  millimeters, depending on application \cite{lockwood1998real}. This prohibits the use of traditional US imaging in applications where the features being imaged are of microscopic scale, such as the microscopic blood vessels developing around malignant cancerous tumors, or around inflamed regions such as the abdomen in Crohn's disease. Adressing this challenge may lead to unprecedented, non-invasive, non-radiative diagnostic capabilities for early detection and/or treatment of many newly detectable pathologies.

This challenge is addressed by reviewing techniques in the field of Contrast-Enhanced UltraSound (CEUS) imaging, where the injection of designated agents into the blood stream allows for the exploitation of prior knowledge about their frequency domain behaviour, as well as their sparsity in several domains \cite{christensen2020super}. This results in an ability to map sub-wavelength features such as the microscopic vessels (micro-vasculature) created near malignant tumors. We review the use of sparsity and deep learning in this context, and demonstrate the improved resolution that can be obtained by algorithms utilizing them, such as the SUSHI method \cite{bar2018sushi}.

We conclude our review by discussing several outstanding challenges related to signal processing and image processing, and the exciting benefits their solution may enable. We hope this review will inspire further work by the signal processing community to bring ultrasound closer to the patient side and pave the way to high quality and portable ultrasound imaging at the home and primary care. 

\section{Basics of Ultrasound Imaging}
\label{sec:challenges}
We begin by detailing the stages taken to perform ultrasound imaging. The most common mode of US imaging is referred to as Brightness mode, or B-mode. In B-mode, a gray-scale image of a desired plane (or volume in the case of 3D) is generated, with the brightness at each point representing its acoustic reflectivity. Scanning is performed using a designated system with a probe (transducer) moved around to determine the imaging region. 

The typical stages of B-mode formation include: Transmission, where the imaging plane is insonified by ultrasonic pulses emitted by the probe; Reception, where their echoes are received in the transducer elements; Delay, where geometric calculations are used to match the timing of the recorded signals across the elements; and Sum, where the signals of all elements are summed (or averaged) to yield a brightness value at each point. The Delay and Sum stages are often lumped together in an algorithm termed Delay-and-Sum (DAS) Beamforming.

Once summed, the signal is referred to as RF data and undergoes further processing to enhance the resulting image, such as envelope detection, up-sampling, dynamic range compression, contrast adjustment and edge enhancement
\cite{hedrick1989image}, before being stored as digital image data. In most commercially available US systems, the sheer size of data before summing usually means that only RF data, or even image data, is exported digitally, with prior stages performed in hardware.

In transmit stage, the imaging region is insonified with acoustic energy, most commonly by the formations of thin, focused, axial beams, transmitted one at a time and swept laterally across the imaging region. In receive stage, the signals acquired from each such beam are referred to as an image `line', since they are each used to form the brightness along a line in the beam axis. To mitigate the effect of echoes from one beam to the reception of another, a beam is  transmitted only after the previous line's echoes are acquired.
Once a line is received, its data may be described as a set of signals $\varphi_m(t;\theta)$, where $m$ denotes the channel number, $t$ denotes time and $\theta$ the lateral coordinate of the beam. In this form, the data is referred to as channel data.

In the delay stage, an appropriate time delay is applied to each channel so that the signals are aligned in time to yield the delayed signal
\begin{equation}
\label{eq:delayed signal in time}
\hat{\varphi}_m(t,\theta) = \varphi_m(\tau_m(t;\theta)),
\end{equation}
where $\hat{\varphi_m}(t,\theta)$ is the delayed signal for channel $m$ at time $t$ and $\tau_m(t;\theta)$ is the time delay for the same channel, given by

\begin{equation}
\label{tau}
\tau_m(t,\theta) = \dfrac{1}{2}(t+\sqrt{t^2-4(\delta_m/c)t\sin \theta +4(\delta_m/c)^2}).
\end{equation}
Here $c=1540 m/sec$ is the effective speed of sound in tissue and $\delta_m$ is the distance between the beam origin and receiving element $m$. In the Sum stage, the signals are summed to yield the brightness-mode value at each point in the line:

\begin{equation}
\label{eq:beamformed signal in time}
\Phi(t;\theta)=\sum_{m}\varphi_m(\tau_m(t;\theta)),
\end{equation}
where $\Phi(t;\theta)$ denotes the beamformed data. Repeating the process for each line yields a brightness value for each point in the imaging region. In practice, the delay is performed in the digital domain after sampling on a sufficiently dense grid, to enable implementation of \eqref{tau} in digital. 

Many improvements upon this basic DAS scheme have been developed over the years, such as apodization and adaptive beamforming. See box "Advanced Beamforming" for more information. In practice, DAS remains the most commonly used beamforming technique, thanks to its simplicity and applicability for real-time imaging.

\begin{figure*}[t!]
    \centering
    \includegraphics[width=0.7\textwidth, height=7.5cm]{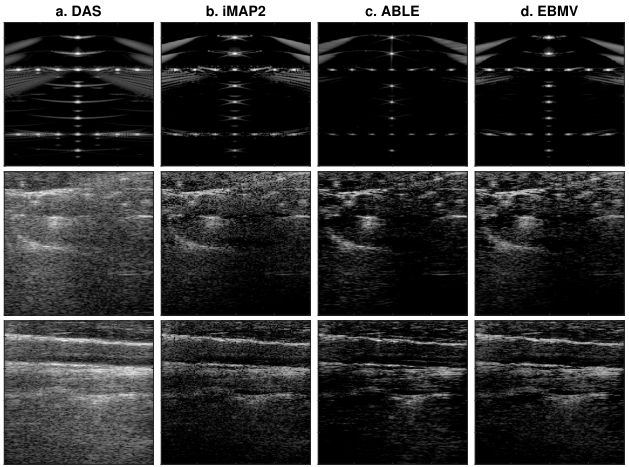}
    \caption{Comparison of adaptive and learned beamforming methods. Single-plane-wave \emph{in-vivo} images using: a) Delay-and-sum (DAS) beamforming with Hanning apodization, b) iMAP2 beamforming, c) adaptive beamforming by deep learning (ABLE), and d) Eigen-Based Minimum variance (EBMV) beamforming. From top to bottom the images are of simulated point scatterers, carotid artery cross-section, and a carotid artery longitudinal cross-section. Figure adapted from \cite{luijten2020adaptive}.}
    \label{fig:adaptive_weighting}
\end{figure*}

Although the method of focused-beam, line-based beamforming described above is the most common in US imaging, other techniques have been developed for various settings. Plane wave and diverging wave schemes insonify the entire scan region with a single pulse allowing for high frame rate ~\cite{matrone2016ultrasound,moghimirad2019plane}. M-Mode imaging narrows the scan region to a single line, monitoring axial movements at a high refresh rate \cite{kumon2012high}. Finally, various versions of Doppler ultrasound use designated transmit schemes to improve spectral resolution and estimate flow velocities \cite{hoskins1990measurement}.

\section{Data-Rate Reduction}
\label{sec:SubNyquist}

As previously described, the standard approach to digital beamforming is through DAS, thanks to its relatively low computational load. The delays applied in DAS typically are on the order of nanoseconds, which results in high sampling rate requirements \cite{demuth1977frequency}. For practical reasons, US signals are sampled at lower rates, in the order of tens of megahertz, with rounding or interpolation used to perform fine delaying. This reduced rate is still higher than the actual Nyquist rate of the signal, which is twice its bandwidth \cite{eldar2015sampling}. For fine delay resolution, a common rule of thumb is to sample at a rate 4-10 times higher than the central frequency of the transducer (which typically is 1-20 MHz, depending on application). Sampling at such high rates leads to huge amounts of data that need to be stored and processed.

Here we review techniques for reduction in sampling rate, based on the combination of compressed sensing (CS) ~\cite{tur2011innovation,chernyakova2013compressed} and sub-Nyquist sampling \cite{eldar2015sampling} and by exploiting the finite rate of innovation (FRI) \cite{vetterli2002sampling} structure of the beamformed signal. We also consider methods based on algorithm unfolding in the recovery stage ~\cite{monga2021algorithm, mamistvalov2021deep}. We begin by describing an approach of equivalently implementing DAS beamforming in the Fourier domain, referred to as frequency domain beamforming (FDBF) \cite{chernyakova2013compressed}. We then show how FDBF can pave the way to sub-Nyquist sampling, leading to rates that are much lower than those imposed by time-domain considerations. The suggested sub-Nyquist system is implemented based on the ideas of Xampling presented in \cite{tur2011innovation} and further elaborated on in the box ``Xampling for US imaging".


\begin{figure*}
\fboxsep1em
\colorbox{BoxBackground}{\begin{minipage}{1\textwidth}\begin{multicols*}{2}
\section*{Xampling for US imaging}
Xampling is a sampling method designed to sample signals below their Nyquist rate, in a way that enables perfect reconstruction \cite{mishali2011xampling}. 
In the context of FRI signals, which is the case in US imaging, Xampling can be used to obtain a sub-Nyquist representation by sampling only a small subset of frequency values. In US imaging, each signal has one main band of energy in the frequency domain, around the central frequency of the probe. Using the Xampling mechanism a subset of the main band of energy is obtained. 

Specifically, let $\kappa$ be an arbitrary set comprised of $K$ consecutive frequency components, and let $\varphi_m(t)$ be the received signal at channel $m$. 
The set of subsamples of the received signal is obtained using an appropriate sampling filter \cite{tur2011innovation}
\begin{equation}
\varphi_m(n_s) = \int_{-\infty}^{\infty} \varphi_m(t)f_r(t-n_sT_{sN})dt,
\end{equation}
where $\varphi_m(n_s)$ is the $n_s$th sample of the analog US signal and
\begin{equation}
f_r(t)  = \sum_{l=-r}^{r}f(t+lT),
\end{equation}
where $r$ is a constant determined by the support of the transmitted pulse, $T$ is the duration of the received signals, and $T_{sN}$ is the sub-Nyquist sampling period.
The filter $f(t)$, is designed according to the required set $\kappa$, and effectively zeros out the frequency components that are not included in $\kappa$. Explicitly, the frequency response of $f(t)$ satisfies
\begin{align}
\label{eq:filter}
F(\omega) = 
\begin{cases}
0, \: \text{if} \:\: \omega = \dfrac{2\pi k}{T},\;  k\notin \kappa \\
1, \: \text{if} \:\: \omega = \dfrac{2\pi k}{T}, \; k\in \kappa \\
$arbitrary$, \:\: \text{else}.
\end{cases}
\end{align}
Therefore, the number of samples is equal to the number of Fourier coefficients of interest. 

After filtering the signal, it is sampled at its effective Nyquist rate.
Finally, applying a Fourier transform on the sampled signal results in the Fourier coefficients of the US signal, enabling the application of Fourier domain beamforming \cite{chernyakova2013compressed}. \ref{Fig:Xampling_US} illustrates Xampling for US imaging.

    \includegraphics[width=1\columnwidth,height=1.7cm]{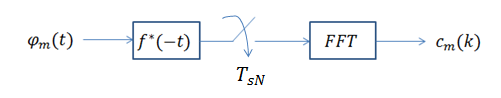}
    \caption{Xampling framework for Fourier domain beamforming. In the first stage, signals received at all channels are filtered using an analog filter that obtains the desired set of frequency components $\kappa$, as described in \eqref{eq:filter}. The signals are then sampled at their effective Nyquist rate, followed by an FFT and application of FDBF, resulting in low rate Fourier coefficients of the beamformed signal.}
    \label{Fig:Xampling_US}
\end{multicols*}
\end{minipage}}\end{figure*}

\subsection{Frequency-Domain Beamforming}
\label{sec:frequency domain beamforming}
In \cite{chernyakova2013compressed} the authors suggested transforming the process of beamforming into the frequency domain in order to reduce the sampling rate, showing that beamforming can be performed in frequency, using a small set of Fourier coefficients. Let $c[k]$ denote the $k$th Fourier series coefficient of the beamformed signal in \eqref{eq:beamformed signal in time} and $\hat{c}_m[k]$ be the $k$th Fourier series coefficient of the delayed signal at channel $m$. Due to the linearity of the Fourier transform and following \eqref{eq:beamformed signal in time},
\begin{equation}\label{FDBF_eq_delay_BF}
c[k] = \dfrac{1}{M} \sum^{M}_{m=1} \hat{c}_m[k],
\end{equation}
where $M$ is the number of channels and $\hat{c}_m[k]$ are the delayed Fourier coefficients, given by
\begin{equation}
\hat{c}_m[k] = \dfrac{1}{T}\int_{0}^{T} I_{[0,T_B(\theta))}(t)\hat{\varphi}_m(t;\theta)e^{\frac{-2\pi j}{T}kt }dt.
\end{equation}
Here, $I_{[a,b)}$ is the indicator function, equal to $1$ when $a\leq t < b$ and zero otherwise. The beam is supported on $\left[ 0, T_B(\theta)\right)$, where 
$T_B(\theta) = \min\limits_{m}{\tau_m^{-1}(t,\theta)} $ and $T$ is the pulse penetration depth \cite{chernyakova2013compressed}.

\begin{figure}[ht]
\centering
\includegraphics[width=8cm]{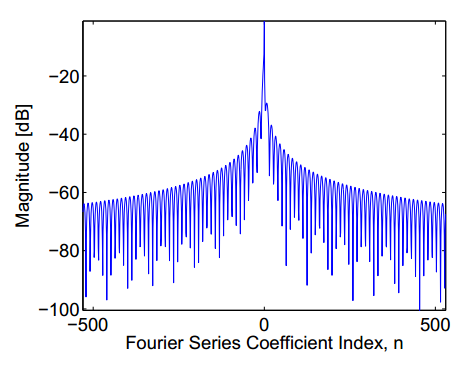}
\caption{Fourier coefficients $Q_{k,m;\theta}[n]$, for Fourier coefficient $k=100$, receiving channel $m=14$, and radiated direction $\theta=0.421 \; rad$. Image adapted from \cite{chernyakova2013compressed}.}
\label{fig:Q}
\end{figure}

In \cite{chernyakova2013compressed}, it is shown that the Fourier coefficients of the delayed signal can be expressed as
\begin{equation} \label{eq_dist_fun_Fourier}
\hat{c}_m[k] = \sum^{\infty}_{n = -\infty} c_m[k-n]Q_{k,m;\theta}[n],
\end{equation}
where $c_m[k]$ are the Fourier coefficients of the received signals in each channel before the delay is applied. The coefficients $Q_{k,m;\theta}[n]$, illustrated in Fig. \ref{fig:Q}, are the Fourier coefficients of a distortion function, determined solely by the the imaging setup and geometry, essentially transforming the delay stage of beamforming from the temporal to the frequency domain. These coefficients may be computed offline and stored in memory for real time applications. 

Due to the decay properties of $\left\lbrace Q_{k,m;\theta}[n]\right\rbrace $, 
the terms in the last sum may be neglected outside a finite set around $n=0$. The summation in \eqref{eq_dist_fun_Fourier} can thus be approximated by a relatively small finite sum 
\begin{equation} \label{eq_dist_fun_Fourier_notFinal_finite}
\hat{c}_m[k] = \sum_{n=-N_1}^{N_2} c_m[k-n]Q_{k,m;\theta}[n]
\end{equation}
where $N_1, N_2$ are chosen empirically. The combination of \eqref{eq_dist_fun_Fourier_notFinal_finite} and \eqref{FDBF_eq_delay_BF} yields the FDBF formula:
\begin{equation} \label{eq_dist_fun_Fourier_finite}
c[k] = \dfrac{1}{M} \sum^{M}_{m=1} \sum_{n=-N_1}^{N_2} c_m[k-n]Q_{k,m;\theta}[n].
\end{equation}
By applying an inverse Fourier transform on $c[k]$, the  beamformed signal is obtained.

The main advantage of this method is that it requires only a limited set of significant Fourier coefficients of the received signals. Each such coefficient is calculated using FFT, as a linear combination of samples of the signal. Thus, for fewer Fourier coefficients, less samples are required, and a lower sampling rate may be used.

\subsection{CS Based Beam Reconstruction}
\label{ssec:beam reconstruction}
A further reduction in sampling rate may be obtained by acquiring only a subset of the non-negligible Fourier coefficients. However, since in this case the signal is sampled below its effective Nyquist rate, applying an inverse FFT results in aliasing in the time domain. To overcome aliasing effects, we exploit the structure of the beamformed signal, relying on techniques from compressed sensing (CS) and finite rate of innovation (FRI) sampling ~\cite{tur2011innovation, eldar2015sampling,chernyakova2013compressed}. More specifically, due to the way a beamformed signal is formed, it can be expressed as a shifted and scaled sum of replicas of the (known) transmitted pulse

\begin{equation}\label{beam FRI}
    \Phi(t;\theta)\simeq\sum_{l=1}^L\tilde{b}_l h(t-t_l),
\end{equation}
where $h(t)$ is the transmitted pulse,  $L$ is the number of scattering elements in direction $\theta$, $\{\tilde{b}_l\}_{l=1}^L$ are the unknown amplitudes of the reflections and $\{t_l\}_{l=1}^L$ denote the time of arrival of the reflection from the $l$th element. Such a signal is completely defined by the $2L$ unknown amplitudes and delays.

By quantizing the delays $\{t_l\}_{l=1}^L$ with quantization step $T_s=\frac{1}{f_s}$, such that $t_l=q_lT_s, q_l\in\mathbb{Z}$, where $T_s$ is the sampling period and $f_s$ is the sampling frequency, we may write the Fourier coefficients of the beamformed signal as:
\begin{align}
\label{eq:Fourier BF FRI}
c[k]=h[k]\sum_{l=0}^{N-1} b_l e^{-i \frac{2\pi}{N}k l},
\end{align}
where $N=\lfloor T/T_s \rfloor$, $h[k]$ are the Fourier coefficients of the transmitted pulse, and
\begin{align}
\label{eq:tilde b - b}
b_l = \left\{ \begin{array}{rl}
 &\hspace{-0.5cm}\tilde{b}_l, ~\mbox{ if $l=q_l$} \\
 &\hspace{-0.5cm} 0, ~~\mbox{ otherwise}.
       \end{array} \right.
\end{align}
%
Defining an $M$-length measurement vector $\mathbf{c}$ with $k$th entry $c[k]$, 
\eqref{eq:Fourier BF FRI} may be rewritten as
\begin{equation}\label{eq:c=HDb}
\mathbf{c}=\mathbf{H} \mathbf{D} \mathbf{b} = \mathbf{A} \mathbf{b} ,
\end{equation}
where $\mathbf{H}$ is an $M\times M$ diagonal matrix with $h[k]$ as its entries, $\mathbf{D}$ is an $M\times N$ matrix formed by taking a set 
of rows from an $N\times N$ Fourier matrix, and $\mathbf{b}$ is a vector of length $N_{st}$ with $l$th entry $b_l$, where $N_{st}$ is the number of samples required for standard DAS beamforming.

Once formulated this way, the problem is that of recovering a sparse vector $\mathbf{b}$, given measurements $\mathbf{c}$.
A typical beamformed ultrasound signal is comprised of a relatively small number of strong reflections and many scattered echoes. Thus, the vector $\mathbf{b}$ defined in \eqref{eq:c=HDb}, is typically not strictly sparse, but rather compressible. This property can be captured by using the $\textit{l}_1$ norm as an objective function in an optimization problem:
\begin{equation}\label{eq:l1 minimization}
    \min_{\mathbf{b}}\|\mathbf{b}\|_1  \textrm{~~~subject to~~~} \|\mathbf{A}\mathbf{b}-\mathbf{c}\|_2\leq\epsilon,
\end{equation}
with $\epsilon$ a parameter representing a noise level. Problem \eqref{eq:l1 minimization} can be solved using second-order methods such as interior point methods \cite{candes2007l1}, 
or first-order methods, based on iterative shrinkage ideas  
such as the NESTA algorithm \cite{chernyakova2013compressed}, which was shown to be highly suitable for US signals. 


\begin{figure*}[ht!]
\fboxsep1em
\colorbox{BoxBackground}{
\begin{minipage}{1\textwidth}
\begin{multicols*}{2}
\section*{Experimental BF demonstration}\label{sec:experimental_demo}
A low-rate frequency domain beamforming mechanism is implemented on an ultrasound imaging system. The setup is shown in Fig. \ref{fig:labsetup}. A state of the art GE ultrasound machine, phantom and ultrasound probe with 64 acquisition channels were used for the scan. The radiated depth $r=15.7$ cm and speed of sound $c=1540$ m/sec yield a signal of duration $T=2r/c\simeq204$ $\mu$sec. The acquired signal is characterized by a narrow band-pass bandwidth of $1.77$ MHz, centered at a carrier frequency $f_0\approx3.4$ MHz. The signals are sampled at the rate of $50$ MHz and are then digitally demodulated and down-sampled to the demodulated processing rate of $f_p\approx2.94$ MHz, resulting in $1224$ samples per transducer element. Linear interpolation is applied in order to improve beamforming resolution, leading to $2448$ samples used to perform beamforming in time.
Fig. \ref{fig:TimeFreqBF}(Left) presents a schematic block diagram of the system's front end.

As illustrated in Fig. \ref{fig:TimeFreqBF}(Right), in-phase and quadrature components of the received signals were used to obtain the desired set of their Fourier coefficients. Using this set, beamforming in frequency was performed according to \eqref{eq_dist_fun_Fourier_finite}, yielding the Fourier coefficients of the beamformed signal. Frequency domain beamforming was performed at a low rate: $M=100$ Fourier coefficients of the beamformed signal were calculated, using $K=120$ Fourier coefficients of each one of the received signals. This corresponds to $240$ real-valued samples used for beamforming in frequency. Hence, beamforming in frequency is performed at a rate corresponding to $240/2448\approx1/10$ of the demodulated processing rate.

This implementation was done on a commercial system which samples each channel at a high rate. Data and processing rate reduction took place following the Fourier transform, in the frequency domain. The set of $120$ Fourier coefficients of the received signals, required for frequency domain beamforming, are obtained from only $120$ low-rate samples.

\begin{center}
    \frame{\includegraphics[width=1\columnwidth,height=3cm]{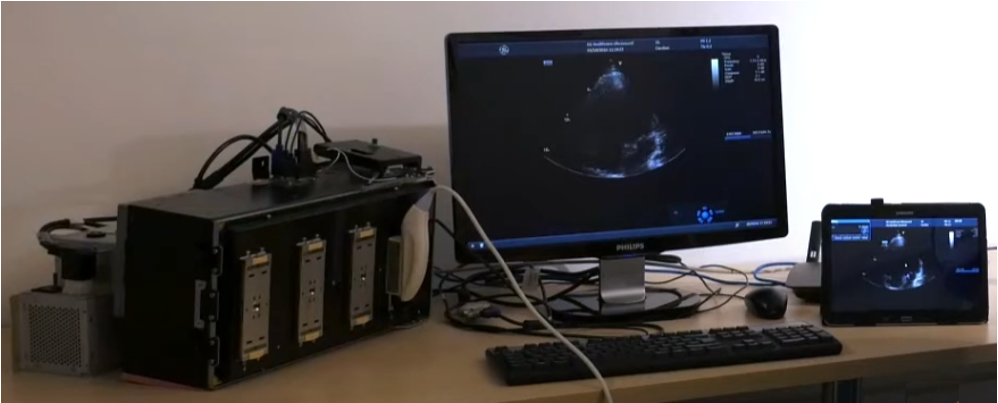}}
    \caption{Lab setup: Research ultrasound system during cardiac scan, with tablet and monitor display.}
    \label{fig:labsetup}
\end{center}

The performance of the proposed method was verified by scanning a healthy consenting volunteer.
Fig. \ref{fig:demoResults} compares low-rate beamforming in frequency and standard time-domain beamforming of a cardiac scan of the volunteer. 
As can be seen, sufficient image quality is retained, despite the significant reduction in processing rate.

\begin{center}
    \includegraphics[width=1\columnwidth]{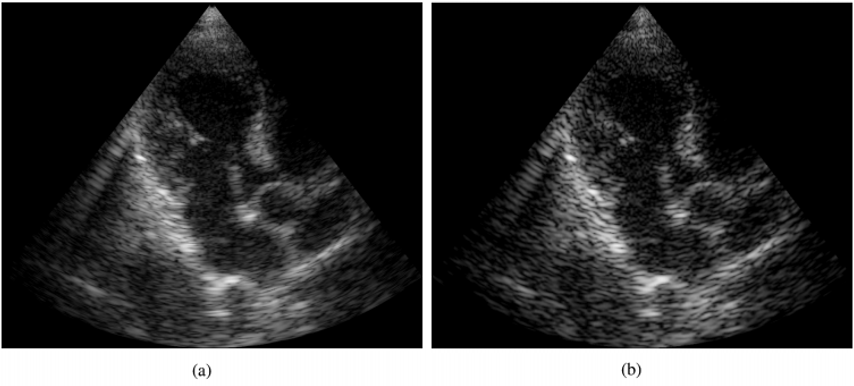}
    \caption{Cardiac imaging. (left) Time domain beamforming. (Right) Frequency domain beamforming, obtained with 28-fold reduction in the processing rate. Image adapted from \cite{chernyakova2013compressed}.}
    \label{fig:demoResults}
    \end{center}

\end{multicols*}

\begin{center}
	\centering
	\includegraphics[width=\textwidth, height=3cm]{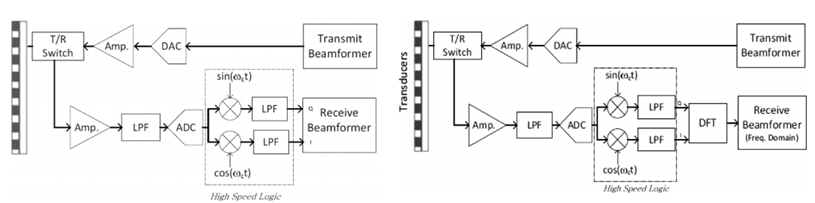}
	\caption{Time domain and frequency domain BF. Image adapted from \cite{chernyakova2013compressed}. (Left) Standard time-domain transmit and receive front-end of a medical ultrasound system. (Right) Transmit and receive paths of a medical ultrasound system with beamforming in the frequency domain.}
	\label{fig:TimeFreqBF}
\end{center}

\end{minipage}}
\end{figure*}

\begin{figure*}[ht]
  \centering
  {\includegraphics[width=0.8\textwidth, height=7cm]{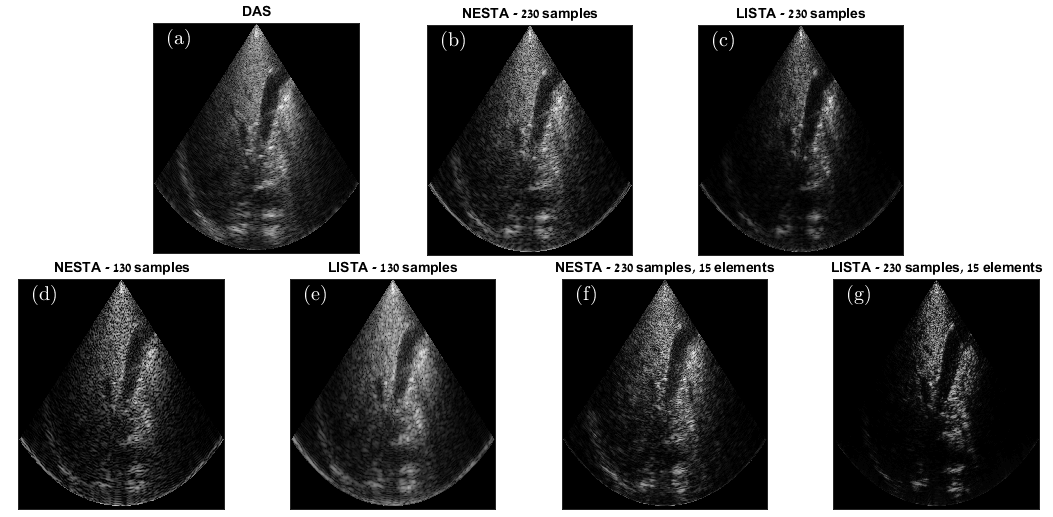}}
\caption{\emph{In-vivo} liver images produced with: (a) DAS - 1920 samples, (b) FDBF + NESTA recovery - 230 samples, 8-fold reduction, (c) FDBF + LISTA recovery - 230 samples, 8-fold reduction, (d) FDBF + NESTA recovery - 130 samples, 15-fold reduction, (e) FDBF + LISTA recovery - 130 samples, 15-fold reduction, (f) CFCOBA + NESTA recovery - 230 samples, 15 channels, 36-fold reduction, (g) CFCOBA + LISTA recovery - 230 samples, 15 channels, 36-fold reduction. Image adapted from \cite{mamistvalov2021deep}.}.
\label{fig:liver_res}
\end{figure*}

\begin{figure}
    \centering
    \hspace{-0.2cm}
    \begin{minipage}[h!]{.4\linewidth}
        \centering
        \includegraphics[width = 0.8\linewidth,height=2.5cm]{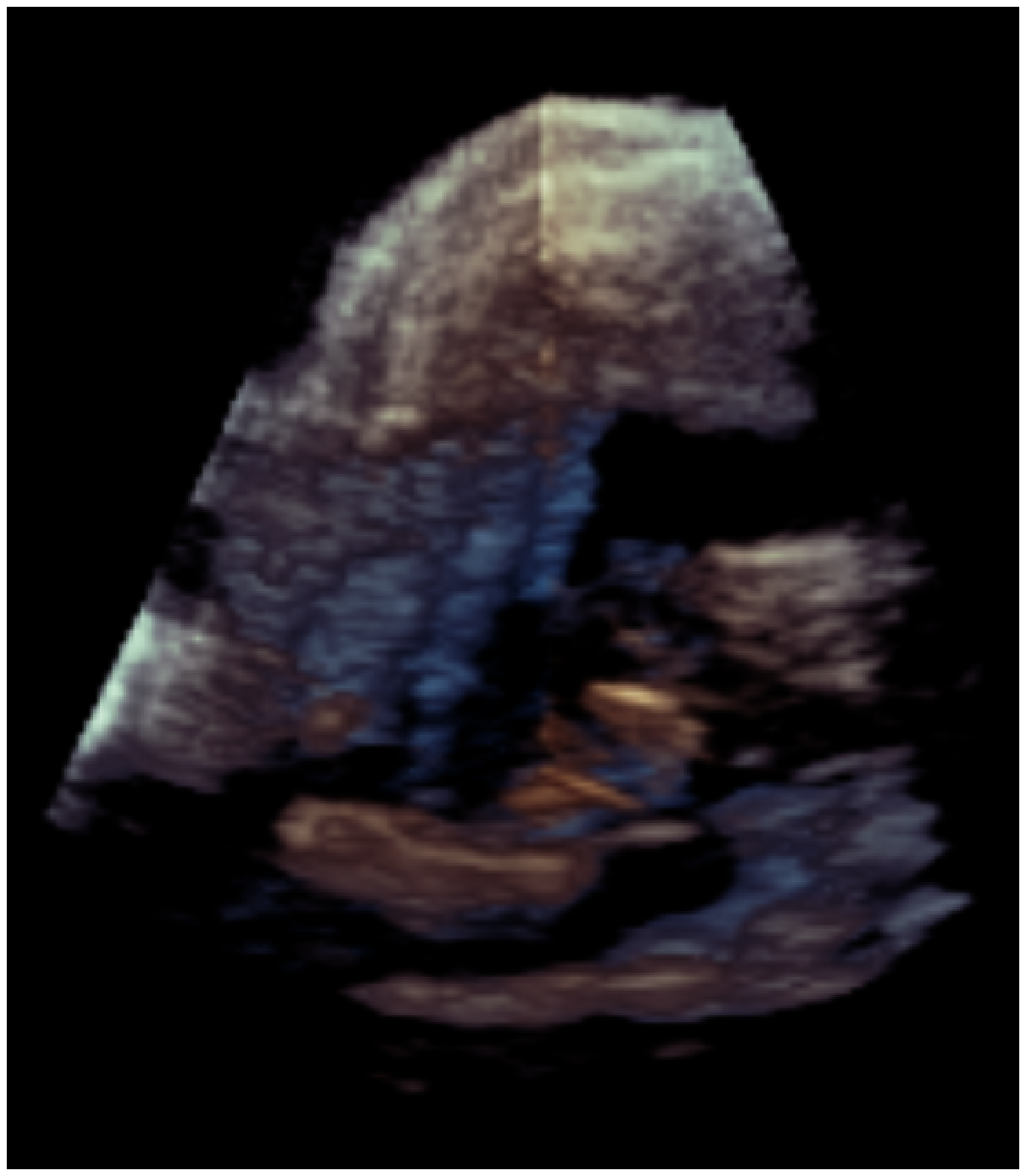}
        \label{subfig:comm time frame}
    \end{minipage}
    \begin{minipage}[ht!]{.4\linewidth}
        \centering
        \includegraphics[width = 0.8\linewidth, height=2.5cm]{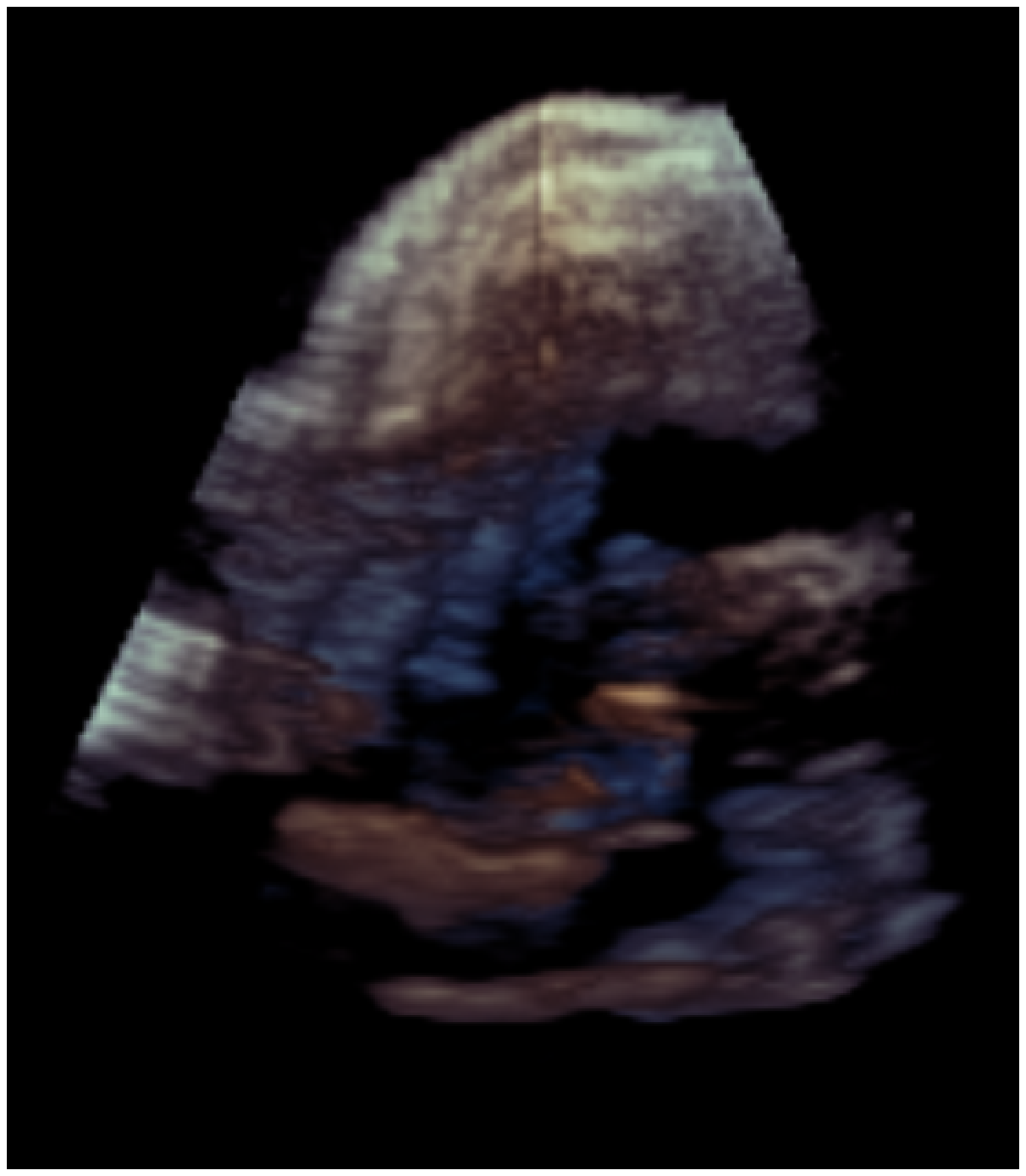}
        \label{subfig:comm half freq frame}
    \end{minipage}
    \caption{3D imaging of a heart ventricle phantom. (Left) Time-domain reconstruction of a single frame. (Right) Frequency-domain reconstructed frame, with 12-fold rate reduction. Image adapted from \cite{burshtein2016sub}.}
    \label{fig:comm volumetric}
\end{figure}

The techniques discussed above, are evaluated in Fig. \ref{fig:liver_res} using \emph{in-vivo} liver data of a healthy volunteer. Acquisition was performed using the Verasonics Vantage 256 System, using the 64-elements phased array transducer P4-2v. The frequency response of this probe is centered at 2.72 MHz, and a sampling rate of 10.8 MHz was used, leading to 1920 samples per image line. In (a), DAS beamformed US image is presented, using 1920 samples. Sub Figs. (b) and (d) show the potential usage of FDBF and CS reconstruction using 230 and 130 samples, leading to 8, and 15 fold reduction in data size and sampling rate, respectively.
A description of a specific imaging setup that was used to implement the suggested methods above is given in the “Experimental BF demonstration” box.
The described methods can also be extended to 3D US imaging \cite{burshtein2016sub}; an example of implementing these methods for 3D imaging is given in Fig. \ref{fig:comm volumetric}, presenting a 12-fold rate reduction.

Another powerful means of solving the recovery problem in real time is through the use of deep learning networks, specifically using algorithm unrolling ~\cite{gregor2010learning, li2020efficient, monga2021algorithm}.
In \cite{mamistvalov2021deep}, it was suggested to use the unrolled LISTA architecture, which is a learned version of the well known ISTA method for sparse recovery. For further details, see ``Sparse Recovery Using LISTA" box. 
As opposed to complicated deep learning approaches for beamforming, LISTA consists of fast and efficient training and testing steps since only a small fraction
of the samples of the beamformed signal needs to be stored for training the network and for full image recovery. In addition, the computational complexity of LISTA, reduces to only $O(n)$, when compared to standard iterative approaches like NESTA, which suffers from a $O(n^2)$ computational load. In Fig. \ref{fig:liver_res}, the usage of LISTA for recovery of subsampled data is depicted in subfigures (c), and (e). Using LISTA, only 230, 130 samples are needed and the resulting images yield better visual and quantitative results compared to NESTA \cite{mamistvalov2021deep}.

The results, shown in Fig. \ref{fig:liver_res}, demonstrate that a significant reduction in sampling rate, while fully recovering the US image is possible. Moreover, the usage of deep unfolding methods, such as LISTA, provides an efficient recovery method for subsampled US signals. The following section, suggests another approach for reduction in data rates by reducing the number of receiving channels.

\begin{figure*}[t!]\fboxsep1em
\colorbox{BoxBackground}{\begin{minipage}{1\textwidth}\begin{multicols*}{2}
\section*{Sparse Recovery using LISTA}
Learned ISTA (LISTA) \cite{gregor2010learning, monga2021algorithm} is a data-driven, learned counterpart of the widely used iterative shrinkage/thresholding algorithm (ISTA) \cite{beck2009fast} for sparse recovery.

Consider a sparse recovery problem
\begin{equation}
\label{eq:sparse_rec}
\min_{\boldsymbol{x}} = \frac{1}{2} || \textbf{y}-\textbf{Ax} ||_2^2 + \lambda ||\textbf{x}||_1 
\end{equation}
where $\textbf{y}$ is the measured signal, $\textbf{A}$ is the measurement matrix, both known, $\lambda $ a regularization parameter 
and $\textbf{x}$ is the underlying sparse vector to be recovered. ISTA solves \eqref{eq:sparse_rec} iteratively through 
the iterations
\begin{equation}
\textbf{x}^{k+1} = S_{\lambda}(\textbf{W}_{e} \textbf{y} + \textbf{W}_t \textbf{x}^k ), \; \textbf{x}^0 = 0,
\end{equation}
where $\textbf{x}^{k+1}, \textbf{x}^k$ are subsequent approximations of the solution, $\textbf{W}_{e} = \mu \textbf{A}^T, \textbf{W}_t = \textbf{I}- \mu \textbf{A}^T \textbf{A}$ for some chosen step size parameter $\mu$, and $S_{\lambda}$ is a soft thresholding operator with threshold $\lambda$.

Adapting ISTA to a learning-based framework is performed by unrolling: cascading layers of a deep network, each mimicking an iteration in the iterated algorithm \cite{monga2021algorithm}. In the unfolded version of LISTA, the layers consist of trainable convolutions $\textbf{W}^k_{e}$, $ \textbf{W}^k_t$, and a trainable shrinkage parameter $\lambda_k$, corresponding to the threshold in ISTA, for $1 \leq k \leq K $, where $K$ is the total number of layers. The thresholding operation in LISTA is replaced by a smooth approximation (though other smooth approximations are possible) 
\begin{equation}
S_{\lambda}(\textbf{x}) = \frac{\textbf{x}}{1+e^{-(|\textbf{x}|-\lambda)}},
\end{equation}
with the operations performed element-wise.

Training LISTA is done in a supervised manner, by first recovering a set of sparse codes $\textbf{x}^i, 1 \leq i \leq N$, given a set of measurements $\textbf{y}^i$ and the known measurement matrix ${\bf A}$. This recovery is typically performed using an iterative sparse solver, e.g. ISTA, FISTA, and NESTA ~\cite{beck2009fast,becker2011nesta}.
Training is typically performed by minimizing the mean-squared-error (MSE) between the unrolled network's output, given the measurements $\textbf{y}^i$ and their respective latent codes $\textbf{x}^i$, 
\begin{equation}
  L ( \textbf{W}_{e} ,  \textbf{W}_t , \lambda ) = \frac{1}{N} \sum_{i=1}^{N} ||\hat{\textbf{x}}^i(\textbf{y}^i; \textbf{W}_{e} ,  \textbf{W}_t , \lambda ) - {\textbf{x}}^i||_2^2,
\end{equation}
using standard loss minimization methods, such as stochastic gradient descent. The basic architecture of LISTA is given in Fig. \ref{Fig:LISTA} within the yellow dashed line.

Among the advantages of unrolled networks over their iterative counterparts are fixed computational complexity (a feed-forward network requires only a single pass), which leads to typically much faster execution times, improved performance, and no need to know the matrix $\boldsymbol{A}$.
Fig. \ref{Fig:LISTA} describes the architecture suggested in \cite{mamistvalov2021deep}, where the last layer $G$, transfers the sparse representation of the output, into a beamformed signal. Unrolled networks may also be used to recover the positions of strong reflectors, such as contrast agents in an US frame, given that their positions in space and time are sparse  \cite{van2019deep,solomon2020robust,van2020super}.

\end{multicols*}
\begin{center}
    \includegraphics[width=0.8\columnwidth,height=4cm]{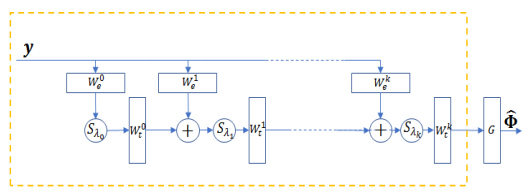}
    \caption{LISTA block diagram, with each layer applying an iteration step (rectangles represent trainable convolutional layers and $S_{\lambda_i}$ represents soft thresholding). $\hat{\Phi}$ stands for the beamformed signal. Image adapted from \cite{mamistvalov2021deep}.}
    \label{Fig:LISTA}
\end{center}
\end{minipage}
}
\end{figure*}

\begin{figure*}[ht]\fboxsep1em
\colorbox{BoxBackground}{
\begin{minipage}
{1\textwidth}
\begin{multicols*}{2}
\section*{Advanced Beamforming}
In US beamforming, channel signals are delayed and summed to yield brightness values in an imaging region, with the objective of spatially focusing and maximizing the energy in the direction of interest. 
Apodization is the process of assigning weights to each channel before summing, to reduce the contribution of certain channels, for example due to low SNR.
Classically, the weights are represented using a window function (e.g. Hamming or Tukey), centered around the element at the center of the beam. 

To formulate apodization mathematically, we consider a ULA of $M$ transducer elements. Assume that all recieved signals have been delayed correctly for a given point in the imaging region. Thus, the delayed signal at channel $m$ can be written as  $y_m = x + n_m$, where $x$ is a signal of interest that represents the reflection of the imaged region and $n_m\sim N(0, \sigma^2_n)$ is additive interference, including off target reflections, clutter and system noise \cite{chernyakova2019imap}. 
In vector form
\begin{equation}
    \boldsymbol{y} = x\boldsymbol{1} + \boldsymbol{n},
\end{equation}
where $\boldsymbol{y}$ is the delayed data, $\boldsymbol{1}$ is a vector of ones and $\boldsymbol{n}$ is the vector of interference.

The classic DAS beamformer, averaging the delayed data over all elements, can be written as
\begin{equation}
    x_{DAS} = \frac{1}{M} \sum_{m=1}^{M} y_m = \frac{1}{M} \boldsymbol{1^H y}.
\end{equation}

With a windowing function $w_m$, we obtain
\begin{equation}
    x_{DAS} = \sum_{m=1}^{M} w_m y_m = \boldsymbol{w^H y}.
\end{equation}

Adaptive beamformers assign weights using the data itself. One such algorithm is the minimum variance (MV) beamformer \cite{vignon2008capon}, which maintains a unity gain in a chosen focusing direction (the foresight), while minimizing the energy received from other directions. It can be written as the solution to the following minimization problem:
\begin{equation}
    \min_{\boldsymbol{w}} E[|\boldsymbol{w^H y}|^2], \; \text{s.t.} \; \boldsymbol{w^H 1} = 1.
\end{equation}
The solution is 
\begin{equation}
    \boldsymbol{w}_{MV} = \dfrac{\boldsymbol{R}_y^{-1}\boldsymbol{1}}{\boldsymbol{1}^H\boldsymbol{R}_y^{-1}\boldsymbol{1}},
\end{equation}
where $\boldsymbol{R}_y = E[\boldsymbol{yy^H}]$ is the received correlation matrix. This requires the inversion of $\boldsymbol{R}_y$, creating computational issues and prohibiting it from being used in practice in US applications. Approaches such as Eigenvalue-Based Minimum Variance, spatial averaging and diagonal loading \cite{luijten2020adaptive} have been developed to remedy this, but they still suffer from a high computational load.

To overcome these issues, Iterative Maximum a-Posteriori (iMAP) beamforming, was suggested in \cite{chernyakova2019imap}. Assuming the beamformed signal is a gaussian random variable $x \sim N(0, \sigma^2_x)$, and that the noise at each element is an uncorrelated Gaussian variable, the iMAP estimator of $x$ is given by
\begin{equation}
    x_{MAP} = \argmax_x p(\boldsymbol{y}|x)p(x).
\end{equation}
The solution is
\begin{equation}
\label{eq: MAP}
    x_{MAP} = \dfrac{\sigma^2_x}{\sigma^2_n + M\sigma^2_x}\boldsymbol{1^H y},
\end{equation}
where the weights
\begin{equation}
    \boldsymbol{w}_{MAP} = \dfrac{\sigma^2_x}{\sigma^2_n + M\sigma^2_x}\boldsymbol{1^H},
\end{equation}
are based on parameters of the prior distribution of the signal of interest and the interference. 

In most realistic scenarios, the parameters are unknown and need to be estimated from the data, hence the use for iMAP. The statistics are calculated based on the previous estimate in an iterative fashion, with $x_{DAS}$ used as the initial estimate:
\begin{equation}
\label{eq: ML}
    \sigma_x^2 = x^2,\; \sigma_n^2 = \frac{1}{M}||\boldsymbol{y} - x\boldsymbol{1}||^2.
\end{equation}
Iterating between \eqref{eq: MAP} and \eqref{eq: ML} yields increasingly better estimates of the distribution of the parameters.

Deep neural networks may also be used to perform adaptive beamforming. One such network, ABLE \cite{luijten2020adaptive}, uses MV beamformed images, computed offline, as targets, to train adaptive weights that minimize the loss function:
\begin{equation}
\label{eq: ML2}
    L = \frac{1}{2}||\log_{10}(\textbf{P}_{ABLE})-\log_{10}(\textbf{P}_{MV})||^2_2
\end{equation}
where $\textbf{P}_{ABLE}$ and $\textbf{P}_{MV}$ are the pixel values of the network output and target, respectively.

A comparison between different adaptive beamforming methods is given in Fig. \ref{fig:adaptive_weighting}, where iMAP2 indicates two iterations of \eqref{eq: MAP} and \eqref{eq: ML}.

\end{multicols*}
\end{minipage}}
\end{figure*}


\section{Sparse array design}
\label{sec:spatSubNyquist}
We now turn to designing arrays which can enable a reduction in the number of required elements, without compromising on image quality. Image contrast and resolution are limited by the number of elements in the probe, leading to designs of hundreds of elements in typical transducers. Thus, techniques that reduce the number of receive channels are of great importance in paving the way to portable low cost devices.

\subsection{Convolutional Beamforming Algorithm (COBA)} \label{sec: COBA}
Recall that \eqref{eq:beamformed signal in time} describes the standard DAS beamforming for appropriately delayed received signals $\varphi_m(\cdot)$, where $m \in M = \{ -(N-1) ... (N-1) \}$, with $M$ being the set of elements comprising the ULA. To create a larger virtual array, it was suggested in \cite{cohen2018sparse} to consider convolutional beamforming, where the beamformed signal is defined as
\begin{equation}\label{Eq:ConvBF}
    \hat{y}(t)=\sum_{n=-(N-1)}^{N-1}\sum_{m=-(N-1)}^{N-1}u_n(t)u_m(t)=\sum_{n=-2(N-1)}^{2(N-1)}s_n(t).
\end{equation}
Here
\begin{equation*}
    s_n(t)=\sum_{i,j:i+j=n}u_i(t)u_j(t),\;n=-2(N-1),\ldots,2(N-1),
\end{equation*}
and 
\begin{equation}
u_m(t)= \exp\{j\angle \varphi_m(t)\}\sqrt{|\varphi_m(t)|},
\end{equation}
with $\angle \varphi_m(t)$ and $|\varphi_m(t)|$ denoting the phase and magnitude of $\varphi_m(t)$, respectively.

Defining $\v s(t)$ and $\v u(t)$ as the vectors of size $2N-1$ whose entries are $s_n(t)$ and $u_n(t)$, respectively, $\v s(t)$ can be written as 
\begin{equation}\label{Eq:convs}
    \v s(t)=\v u(t)\underset{s}*\v u(t),
\end{equation}
where $\underset{s}*$ denotes a linear convolution in the lateral direction, which can be implemented efficiently using the convolution theorem and FFT.
This beamforming procedure is referred to as convolutional beamforming algorithm (COBA).

The advantage of this algorithm may be shown by a comparison between the beam patterns, $H_{DAS}$ and $H_{COBA}$, of DAS and COBA, respectively, assuming a uniform linear array. For DAS, the beam pattern is given by \cite{cohen2018sparse}
\begin{equation} \label{DAS_BP}
H_{DAS}(\theta) = \sum_{n=-(N-1)}^{N-1} \exp\left( {-j\omega_0\frac{\delta \sin\theta}{c}n}\right) ,
\end{equation}
where $\omega_0$ is the central frequency of the transducer and $\delta$ is the distance between two adjacent array elements.
Based on \eqref{Eq:ConvBF}, the beam pattern of COBA can be written as the product of two DAS beam patterns, leading to
\begin{equation} \label{COBA_final_BP}
H_{COBA}(\theta) =  \sum_{n=-2(N-1)}^{2(N-1)} a_n\exp\left( {-j\omega_0\frac{\delta \sin\theta}{c}n}\right),
\end{equation}
where $a_n$ is the intrinsic apodization, calculated by $\boldsymbol{a} = \mathbb{I}_M \underset{s}{*} \mathbb{I}_M$, with $\mathbb{I}_M$ denoting a binary vector whose $m$th entry is 1 if $m \in M$. 
The significance of the last expression is in demonstrating that COBA yields a beam pattern equivalent to a DAS beamformer using a ULA twice as large as the original array used for imaging. 



As \eqref{COBA_final_BP} implies, effectively the array used for imaging is larger, and corresponds to the convolution of the original array with itself. Therefore, \cite{cohen2018sparse} suggested using a sparse array, which is formed by removing some of the elements from the original array such that a desired beam pattern is formed, in a method called sparse COBA (SCOBA). The essential idea is to remove elements such that after convolution, the original beampattern is preserved.

More specifically, the behaviour of the sparse array is determined by the sum co-array \cite{liu2017maximally}. Given an arbitrary array $U$, the sum co-array $S_U$ is the array that includes all distinct elements of the form $n+m$ where $n, m \in U$. The beam pattern of the sparse convolutionally beamformed signal is
\begin{align}\label{Eq:beampat}
    &H_{SCOBA}(\theta) = H_{DAS, U}(\theta)H_{DAS, U}(\theta) \\ \nonumber
    =&\sum_{n,m\in U}e^{-j\omega_0\frac{d\sin(\theta)}{c}(n+m)} 
    = \sum_{l\in S_U}e^{-j\omega_0\frac{d\sin(\theta)}{c}l}, 
\end{align}
where $\theta$ is the angular direction, $c$ is the speed of sound and $\omega_0$ is the transducer central frequency. By appropriately choosing the sparse array $U \subseteq M$ in a way that its sum co-array yields a full ULA, the reconstructed US image after COBA will have the same image quality as DAS beamforming. One can also use sparse COBA for super resolution (SCOBAR), when choosing a sparse array such that its sum co-array is larger than the original ULA, which leads to enhanced image contrast and resolution. Further reduction in the number of elements was considered by using a fractal array geometry \cite{werner2003overview,puente1996fractal,cohen2020sparse}. The box, "The Sum Co-Array" describes several of these array geometries. 

\begin{figure*}[ht!]
\begin{center}
    \centering
    \includegraphics[width=0.8\textwidth, height=6cm]{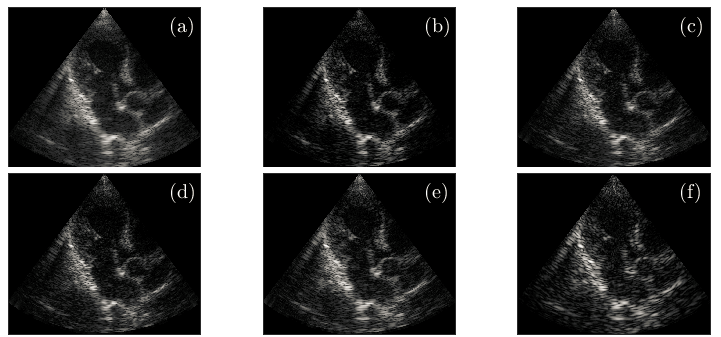}
    \caption{Cardiac ultrasound images beamformed with (a) DAS - 64 elements (ULA), 3328 samples, (b) COBA - full ULA, 3328 samples, (c) SCOBA - 21 elements, 3328 samples, (d) SCOBAR - 27 elements, 3328 samples, (e) fractal geometry - 15 elements, 3328 samples, (f) CFCOBA - 15 elements, 100 samples. The US signals are obtained with different number of receiving channels, and images are produced using COBA, and the compressed frequency domain COBA approach. Images adapted from ~\cite{cohen2018sparse,mamistvalov2020compressed}.}
    \label{Fig:GE_COBA}

\end{center}
\end{figure*}

\begin{figure*}[!ht]

\fboxsep1em
\colorbox{BoxBackground}{
\begin{minipage}{1\textwidth}
\begin{multicols*}{2}
\section*{The Sum Co-Array}\label{sec:sumco}
Assume a uniform linear array (ULA) with $2N-1$ elements, at positions $I=\{-(N-1),\ldots,(N-1)\}$, on a grid with spacing $\delta$. This array can be thinned while exhibiting a fully sampled equivalent array after COBA. This can be utilized for sparse beamforming, without compromising image quality. 

The sum co-array for a ULA with position set $I$ is defined as the set 
\begin{equation}
    \tilde{S}_I=\{n+m:n,m\in I\}.
\end{equation}

This includes the positions of all array elements, as well positions obtained by their summation/difference.

A sparse array with position set $J$ can be obtained from the full ULA with position set $I$ by removing some of its elements. Ideal removal of the elements would be such that they lie on the sum co-array of $J$:
\begin{equation}\label{Eq:subsetJI}
    J\subset I\subseteq S_J.
\end{equation}
In \eqref{Eq:subsetJI}, $I, J$ and $S_j$ are integer sets such that $I$ represents the element locations of the fully sampled ULA, while $J$ and $S_j$ represent the element locations for the thinned array and its resulting sum co-array.

Now, assume that $N$ is not prime and can be factored as $N=AB$, where $A,B=\mathbb{N}^+$ (natural numbers including zero). Define
\begin{equation}\label{Eq:coarray2}
    \begin{array}{ll}
         U_A=\{-(A-1),\ldots,(A-1)\},  \\
         U_B=\{nA:n=-(B-1),\ldots,(B-1)\}
    \end{array}.
\end{equation}
Let $U_A+U_B=\{n+m:n\in U_A, m\in U_B\}$. Then 
\begin{equation} \label{eq:scoba_array}
    U_A+U_B=\{-(AB-1),\ldots,(AB-1)\}=I.
\end{equation}

Denoting $U\subset I$ the array geometry defined as 
\begin{equation}\label{Eq:sumset}
    U=U_A \cup U_B,
\end{equation}
it holds that $I\subset S_U$, where $S_U$ is the sumset of $U$. Therefore, the family of sets \eqref{Eq:sumset} satisfies \eqref{Eq:subsetJI}, and the number of elements in each set is $2A+2B-3$. This configuration does not achieve the set $S_I$ which corresponds to the sum co-array of the full ULA $I$. To achieve this, another set is introduced 
\begin{equation}
    U_C=\{n:|n|=N_A,\ldots,N-1\},
\end{equation}
and now the sparse array geometry is defined as 
\begin{equation}\label{Eq:Varray}
    V=U_A\cup U_B \cup U_C.
\end{equation}
The array $V$ is in fact obtained by adding two small ULAs to $U$, each of size $A-1$ at its edges. It can be verified that 
\begin{equation}\label{Eq:sumeq}
    V\subset I\subset S_V=S_I.
\end{equation}
This implies that elements positioned on the sum co-array of $V$ can produce the same images as those with the sum co-array of the full ULA $I$, despite having fewer elements. 

Further reduction in the size of the array is obtained using fractal arrays \cite{cohen2020sparse}, which are defined recursively by
\begin{align} \label{fractal}
	&W_0 = {0}, \nonumber \\
	&W_{r+1} = \cup_{n\in \mathbb{G}}(W_r + nL^r), \; r\in \mathbb{N},
\end{align}
where $r$ is the array order, $\mathbb{G}$ is the \emph{generator} array in fractal terminology, and $\min(\mathbb{G}) = 0$. The translation factor $L$ is given by $L = 2\max(\mathbb{G})+1$. By choosing the generator array and the fractal order appropriately, desired beampatterns can be obtained.

\end{multicols*}

\begin{center}
    \frame{\includegraphics[width=0.8\columnwidth, height=4cm]{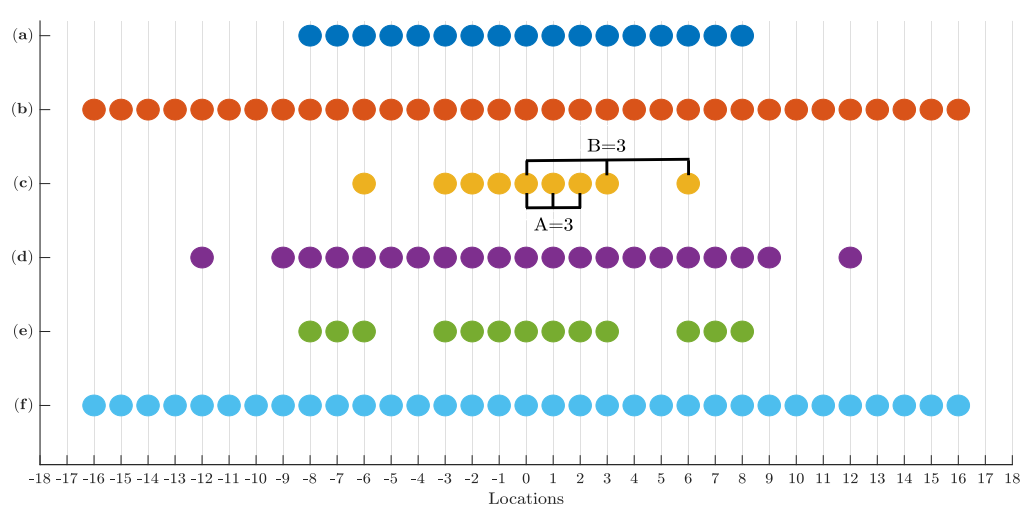}}
    \caption{Element positions of (a) ULA $I=[-8,8]$, (b) sum co-array $S_I=[-16,16]$, (c) sparse array $U$ given by \eqref{Eq:sumset}, (d) sum co-array $S_U$, (e) sparse array $V$ given by \eqref{Eq:Varray} and (f) sum co-array $S_V$. In this example, $d=1,N=9,A=3$ and $B=3$. As can be seen the sum co-arrays of the chosen sparse arrays are bigger or equal to the original ULA, effectively leading to at least the same beam pattern as DAS. Image adapted from \cite{cohen2018sparse}.}
    \label{Fig:sumcoarray}
 \end{center}

\end{minipage}}
\end{figure*}

\ref{Fig:GE_COBA} compares \emph{in-vivo} cardiac image reconstructions of a standard DAS beamformer and the convolutional beamformers presented here, using different array geometries for SCOBA, and SCOBAR. The data was captured using a GE breadboard, phased array probe with 64 channels. The radiated depth was 16 cm, probe carrier frequency 3.4 MHz and sampling frequency 16 MHz. The number of elements used was 21 for SCOBA based on the geometry given by \eqref{eq:scoba_array}, 27 elements for SCOBAR using \eqref{Eq:Varray}, and 15 elements for SCOBA with fractal array geometry of order 3 and generator array $G = \{ 0, 1 \}$. The DAS beamformed image is shown in \ref{Fig:GE_COBA} (a), while (b) demonstrates the improvement of using COBA with the full ULA. SCOBA, and fractal geometry based SCOBA are presented in (c), (e) and achieve similar image quality to DAS with a fourth of the number of elements, while SCOBAR, presented in (d), yields even higher resolution and improved image quality.

\subsection{Combining Low-Rate Sampling and Sparse Arrays}
\begin{figure}[h!]
\centering
\includegraphics[width=1\columnwidth ]{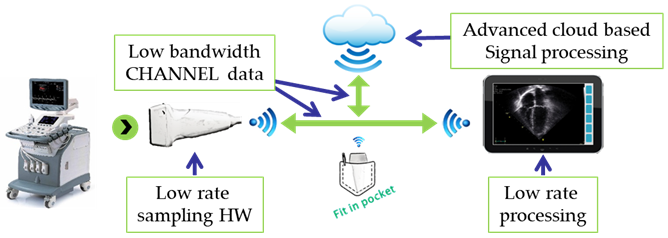}
\caption{Cloud-based US processing (illustration). Probe is connected to the cloud for sampling, on cloud computation, and display tablet. Detaching the processing task from data acquisition will reduce machine size and increase user flexibility, as well as encourage new algorithm development. This also makes data available outside the confines of the specific machine it was acquired with, regardless of its vendor.}
\label{fig:UScloud}
\end{figure}

The ideas of temporal (sampling) and spatial (array design) dilution, discussed in the last two sections, may be combined, achieving further reduction in data size. This technique is referred to as the Compressed Frequency domain Convolutional Beamforming Algorithm (CFCOBA) \cite{mamistvalov2020compressed}.
The idea in CFCOBA, is to calculate the sub-sampled Fourier coefficients of the convolutionally beamformed signal, and to reconstruct it using NESTA or LISTA, based on the FRI and CS frameworks. In \cite{mamistvalov2020compressed}, it is shown that the convolutionally beamformed signal obeys an FRI model, using the square of the known transmitted pulse, hence it can be recovered based on the same approach discussed in Section \ref{ssec:beam reconstruction}. In Fig. \ref{fig:liver_res} (f) the combination of FDBF and COBA is demonstrated, using 230 samples out of 1920 used for DAS and only 15 receiving channels, based on the fractal array geometry, leading to an overall 36-fold reduction in data size for a liver image. Another example of the huge reduction is depicted in Fig. \ref{Fig:GE_COBA} (f), where \emph{in-vivo} cardiac images were acquired with the setup described in Section \ref{sec: COBA}, using a GE breadboard US machine. It can be seen, that using CFCOBA, only 100 samples and 15 elements out of 3328 samples and 64 elements, respectively, are needed for high quality reconstruction, yielding a rate reduction by a factor of 142.
Reconstruction of combined time and space sub sampled data using the LISTA approach ~\cite{monga2021algorithm, mamistvalov2021deep}, as described in Section \ref{ssec:beam reconstruction} and further elaborated on in the ``Sparse Recovery Using LISTA" box, is shown in Fig. \ref{fig:liver_res} (g), which depicts 36-fold reduction in data rates for reconstruction of the beamformed signal using the efficient LISTA algorithm.
This huge reduction in data size, makes wireless US systems possible, as illustrated in Fig. \ref{fig:UScloud}.

\section{Super-Resolution Ultrasound Imaging}
\label{sec:SR}

We now review US imaging performed at a resolution finer than its wavelength, and discuss the methods to obtain it.

As a wave based imaging modality, the resolution of ultrasound imaging is bound to the diffraction limit, meaning that it cannot accurately image objects on a smaller scale than its wavelength. The wavelength used during scanning is often determined by considerations of imaging depth and transducer constraints, and thus cannot be chosen to be arbitrarily small. This practically limits the capabilities of US imaging at roughly a tenth of a millimetre.
 
The vast majority of modern super-resolution ultrasound methods rely on Contrast Enhanced UltraSound (CEUS) ~\cite{dietrich2016perform,solomon2019exploiting}, where an ultrasound contrast agent (UCA), is injected into the bloodstream and then used for localization. The UCA is composed of microscopic bubbles (Microbubbles or MBs). Prior knowledge about the differing behaviour of the MBs and tissue in the scanning region is then exploited in order to separate their respective signals. Processing of the MB signal is performed by relying on further prior knowledge such as their sparsity or shape (as observed in the US image) used to pinpoint locations of MBs over multiple frames. Stacking the frames yields a vascularity map of a finer resolution than achieved in B-Mode imaging.

We first discuss methods relying on localization - the tracking of, and subsequent replacement of MBs with point objects. Then, we show how sparsity in different domains can be exploited to yield non-localization based techniques.
 
\subsection{Ultrasound Localization Microscopy}
\begin{figure}[h!]
    \centering
    \includegraphics[width=1\linewidth]{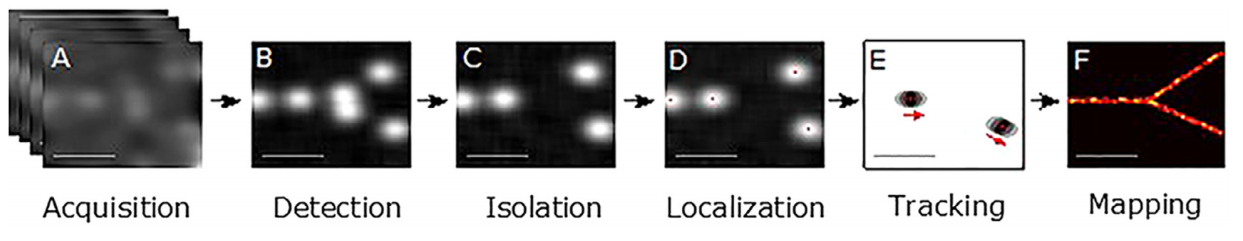}
    \caption{Steps of ULM imaging. (a) Acquisition of ultrasound data over time from contrast enhanced vascular region. (b) Detection of signals from microbubble contrast agents. (c) Isolation of individual microbubble signals; overlapping or interfering signals are rejected. (d) Localization of microbubbles at a precision far beyond the diffraction-limited resolution. (e) Tracking of the microbubbles through consecutive frames to establish velocity profiles. (f) Mapping of the accumulated localizations produces an image of the vascular structure far beyond the diffraction limit. Image adapted from \cite{christensen2020super}.}
    \label{Fig:ULM}
\end{figure}

Ultrasound Localization Microscopy (ULM) is a family of algorithms which achieve super resolution by a simple principle: Objects of sub-wavelength size undergo diffraction, yielding a blurred shape in the US image. Thus, given an observed blurred shape at a certain point in space, it may be replaced by a point source representing a MB. The set of all points where an MB is detected is then used to construct a super resolved vascularity map \cite{christensen2020super}. The stages of ULM consist of: Acquisition, where multiple frames are acquired; Detection, where MB signal is separated from tissue; Isolation, where nearby or overlapping MB signals are discarded;  Localization, where the blurred shape created due to the diffraction of each MB is replaced by a point reflector; Tracking, where estimation of flow is done based on MB positional differences between consecutive frames; and Mapping, where the data is accumulated and post-processed for a visualization of the arterial map. An illustration of these stages is given in \ref{Fig:ULM}.

During acquisition, a large number of frames is acquired. Hundreds, or even thousands of frames may be needed in order to produce a satisfactory vascular map. To this end, ultrafast imaging methods have been developed, often trading off nominal image resolution in favor of frame rate \cite{van2020super}. Using frame rates in excess of 500 frames per second, an analogue to optical localization microscopy is obtained by capturing the transient signal decorrelation of MBs using plane wave imaging \cite{moghimirad2019plane}. Furthermore, a designated transmit-receive scheme may be used in order to enable improved separation of MB from tissue, such as harmonic imaging techniques exploiting their pulsating behaviour; for further details, see ``Harmonic Imaging in Contrast Enhanced Ultrasound" box. 

In the detection stage, the positions of microbubbles within each frame are determined, and they are separated from the tissue. Classically, a centroid localization approach was used, typically by fitting a two-dimensional gaussian brightness curve per MB. More recently, assuming the MBs are sparse i.e. only a few are present per frame, their detection has been treated as a sparse recovery problem \cite{van2020super}. The formulation and example methods for sparse recovery are given in the ``Sparse Recovery Using LISTA" box. For example, one can define $\boldsymbol{x}$ as the (vectorized) map of MB locations and seek a super resolved image by solving for a sparse $\boldsymbol{x}$ using
\begin{equation}
\label{sparserecoveryformula}
    \boldsymbol{x} = \argmin_{\boldsymbol{x}} {||\boldsymbol{Ax - y}||_2^2+\lambda ||\boldsymbol{x}||_1}.
\end{equation}
Here $\boldsymbol{y}$ is the vectorized image frame, and $\boldsymbol{A}$ is the measurement matrix, with each column being a shifted version of the point spread function (PSF), an estimated blurring filter representing the system's response to a point object. Using LISTA to solve \eqref{sparserecoveryformula} enables super resolution without requiring exact knowledge of the system PSF and yields high resolution recovery techniques \cite{van2019deep}.

Despite its high spatial resolution, ULM's reliance on MB sparsity for the sake of detection and isolation implies that MB concentration must be limited during scanning. This, in turn, means that only a small number of MBs may be detected per frame, leading to a requirement for a large number of frames, with long associated acquisition periods, dictating stringent restrictions on patients and forcing frames with significant overlap to be discarded, lowering efficacy.

\begin{figure}[t!]

\begin{center}
    \includegraphics[width=1\columnwidth, height=4cm]{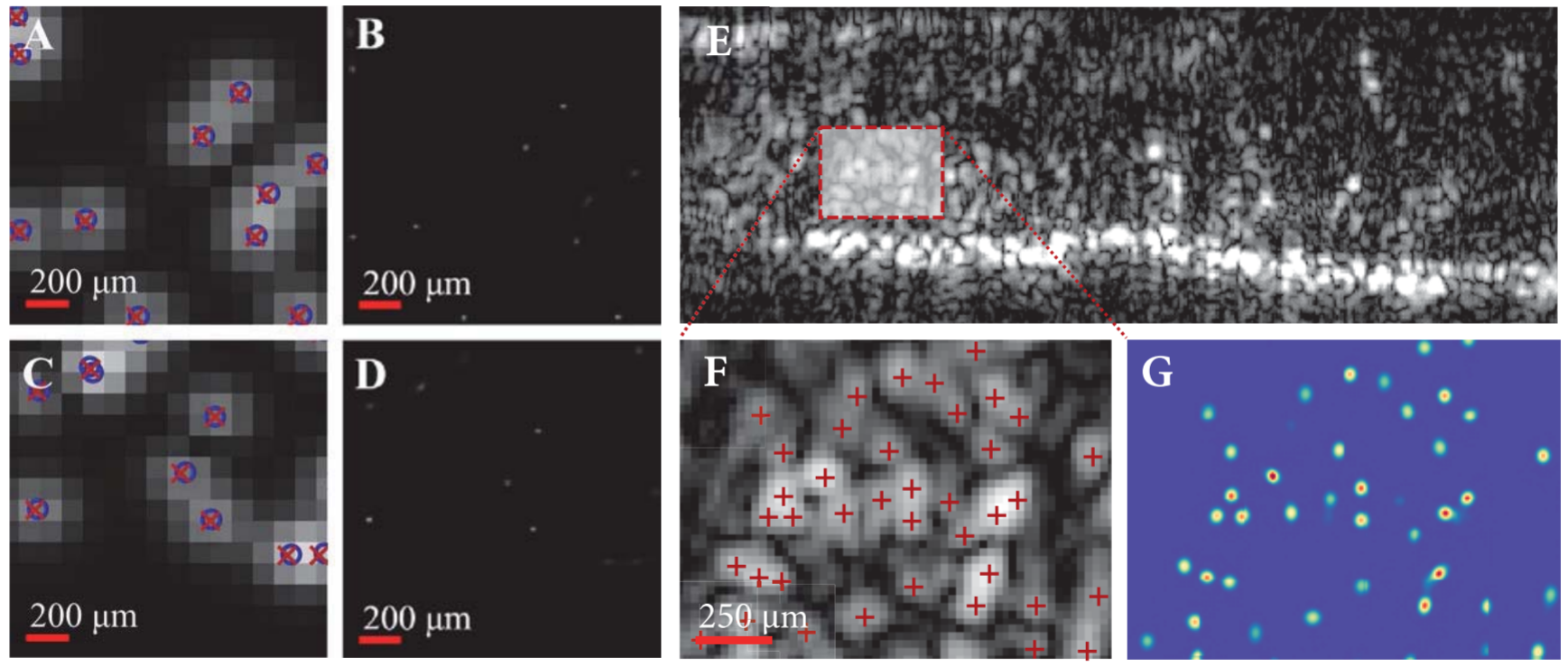}
    \caption{Deep-ULM. (a) and (c) US frames of simulated microbubbles at known positions and point-spread functions estimated from empirical data. (b) and (d) Deep-ULM recoveries for these frames. (e) Example of an in-vivo scan, (f) zoom in on a region of interest in (e), red crosses denote localized MBs, and (g) Deep-ULM recovered MB map. Image adapted from \cite{van2020super}.}.
    \label{Fig:deep_ULM}
\end{center}
\end{figure}

The advent of deep learning in recent years has led to the development of methods that can circumvent the requirement for sparsity by analyzing interference patterns between adjacent MBs. One such example is the frame-based Deep ULM method presented in \cite{van2020super}. Deep ULM generates simulated ground-truth data representing microbubbles at known locations on a grid denser by orders of magnitude than those of the ultrasound data. The MB positions are convolved with point spread functions estimated from empirical measurements, representing the blurred shape of the MBs due to diffraction. The network is then trained to minimize the difference between the ground truth and evaluated positioning of the MBs for different positions and densities. The process is illustrated in \ref{Fig:deep_ULM}.

Other approaches for separating MBs from the tissue include spatio-temporal filtering based on the singular value decomposition (SVD) \cite{van2019deep}. SVD filtering includes collecting a series of consecutive frames, stacking them as vectors in a matrix, performing SVD of the matrix and removing the largest singular values, assumed to be related to the tissue. To overcome difficulties of SVD filtering, the task of clutter removal was formulated as a convex optimization problem by leveraging a low-rank-and-sparse decomposition in \cite{solomon2020robust}. Then, the authors proposed an efficient deep learning solution to this optimization problem through unfolding a robust PCA algorithm, referred to as Convolutional rObust pRincipal
cOmpoNent Analysis (CORONA). This approach harnesses the power of both deep learning and model-based frameworks, and leads to improved separation performance.

\begin{figure*}[h!]
\fboxsep1em
\colorbox{BoxBackground}{
\begin{minipage}{1\textwidth}\begin{multicols*}{2}
\section*{Harmonic Imaging in Contrast Enhanced Ultrasound}

Contrast-enhanced ultrasound imaging is enabled by injection of a contrast agent, consisting of microscopic, gas-filled bubbles, encapsulated in a phospholipid or protein shell \cite{hyvelin2017characteristics}, into the blood stream. These MBs have a typical mean diameter between $1-3\mu m$, 
and circulate in the patient's bloodstream for a few minutes before dissolving. 

MBs exhibit nonlinear vibration when insonified in their resonance frequency, even in low pressure, meaning that their spectrum exhibits higher order harmonics, where linear tissue typically will only reflect the same frequency transmitted by the probe. Tissue signal interferes with the signal received from blood vessels, reducing image contrast. Several schemes for tissue suppression are currently in practice, and depend on two key conditions: Transmitted energy in the MB resonance frequency, and MB motion relative to tissue. Where a transmit frequency close to that of the MB resonance frequency is used, significant sub-harmonics and higher order harmonics are present in the MB signal, and harmonic imaging methods may be used.

Harmonic imaging involves two key methods: Pulse inversion (illustrated in \ref{Fig:SecHarm}), where consecutive pulses have inverse polarity, and amplitude modulation, where the pulses have differing amplitudes. 

In pulse inversion, the sum of received signals cancels the linear (fundamental frequency) part of the signal, whereas in amplitude modulation, the difference between received signals, with scaling to equalize their amplitudes, will do the same. High-pass filtering may also be applied to further reduce residual tissue signal. After processing, the data contains mostly harmonic signals and can be processed in a normal B-mode beamforming (where imaging quality might be improved depending on application) or in a CEUS imaging scheme.

    \includegraphics[width=1\columnwidth, height=3.5cm]{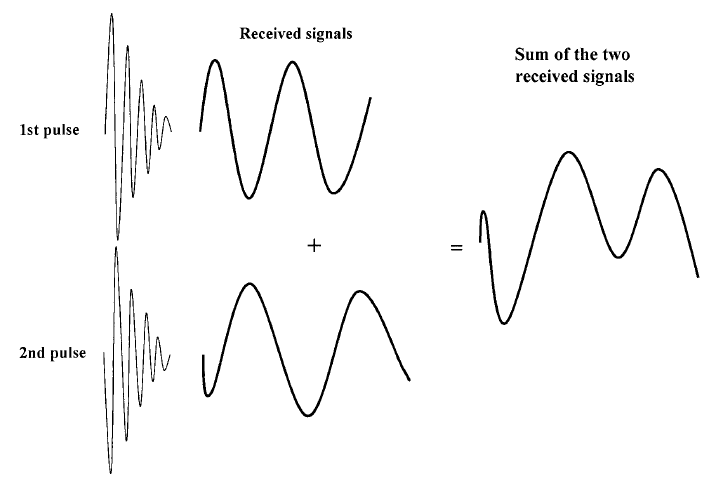}
    \caption{Pulse inversion. For nonlinear media such as UCA's. Inverted pulses do not result in an inverted response, and a residual signal is left following summation, as opposed to linear medium like a tissue, where inverted pulses sum to zero.}
    \label{Fig:SecHarm}

\end{multicols*}
\end{minipage}}
\end{figure*}
\begin{figure}[t!]
    \centering
    \includegraphics[width=0.65\columnwidth,height=2.5cm]{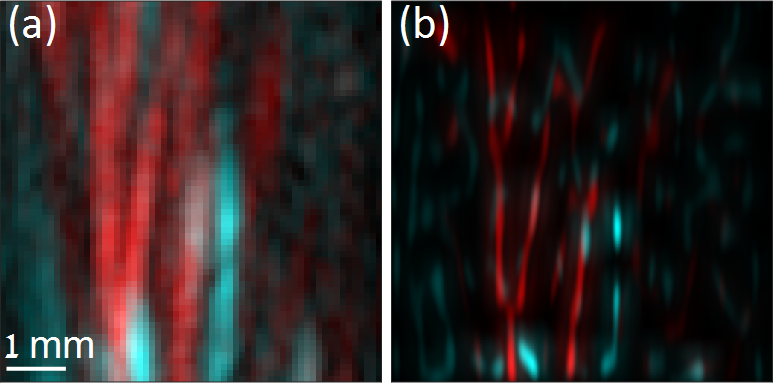}
    \caption{SUSHI performed on a Rabbit's kidney. (a) normally resolved image (b) SUSHI super resolved image. Red and blue coloring represents inward and outward flow components, respectively. Image adapted from \cite{bar2018sushi}.}
    \label{Fig:KidSUSHIp}
 \end{figure}
 \begin{figure}[t!]
\begin{center}
    \centering
    \includegraphics[width=0.75\columnwidth,height=3cm]{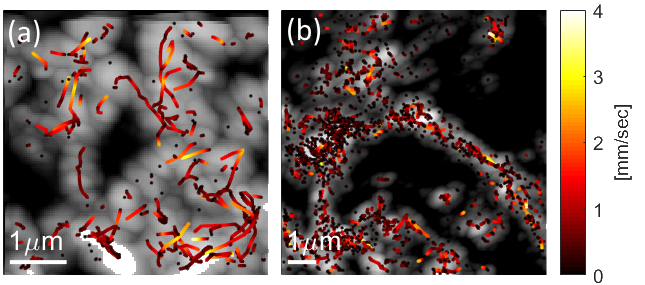}
    \caption{ Super-resolution imaging of two parts of a human prostate using Triple-SAT. (a) normally resolved image with detected MB tracks. (b) super resolved arterial map with MB tracks. Images adapted from \cite{solomon2019exploiting}.}
    \label{Fig:Prost_example}
 \end{center}
\end{figure}

\subsection{Non-localization methods for super-resolution}

Several approaches have been developed to mitigate the reliance of super-resolution imaging on sparsity alone. Here we review two of them: SUSHI, based on exploitation of 2nd-order statistics of a pixel between frames \cite{bar2018sushi}, and Triple-SAT, which exploits flow estimation 
\cite{solomon2019exploiting}.

Sparsity-based super-resolution ultrasound hemodynamic imaging (SUSHI) makes use of the statistical independence between signal fluctuations, over time, of different vessels \cite{bar2018sushi}. This is done through estimation of the per-pixel autocorrelation, which can be expressed as
\begin{align}
    &g_2[m\Delta_{xL},l\Delta_{xL},\tau] = \sum_n |h[m\Delta_{xL} - x_n, l\Delta_{xL}-z_n]|^2 g_n[\tau] + \\ \nonumber
    &\sum_{i,l i \ne l} h[m\Delta_{xL}, l\Delta_{xL}-z_i]\bar{h} [m\Delta_{xL}-x_l, l\Delta_{xL}-z_l]g_{il}[\tau],
\end{align}
where $h$ is the system PSF, $\Delta_{xL}$ is the pixel size, $g_n[\tau]$ is the autocorrelation function of the temporal fluctuations of a pixel $n$ and $g_{il}[\tau]$ is the cross correlation of pixels $i,l$. Here, $\tau$ stands for discrete pre-determined delay of the autocorrelation function, and $i,l$ are indices of dependent volume cells located in the same streamline. The significance of this expression is that MBs flowing independently in different vessels are uncorrelated, and therefore yield only expressions of the form of the first term, where the squaring of the signal creates a squaring of their PSF, making it narrower and thus improving resolution.

A dense imaging grid with microscopic resolution is then created, where the vascular map can be assumed spatially sparse. 
SUSHI exploits this to perform sparse recovery in the spatial domain (e.g. using LISTA). Note that the sparsity assumption holds, even if the microbubbles densely populate a region of the frame, making use of MBs even if they themselves are not sparsely distributed locally. \ref{Fig:KidSUSHIp} demonstrates SUSHI on a Rabbit Kidney, illustrating the blurring caused by diffraction and the resulting, super-resolved arterial map. Due to it's reliance on temporal statistics of the signal, SUSHI's performance depends on the achievable frame rate, and becomes more effective with the advent of ultra-fast imaging modes \cite{bar2018sushi}.
 
When frame rates are low compared with vascular flow velocities the simultaneous, sparsity-based super-resolution and tracking (triple-SAT) may be used. Triple-SAT uses optical flow estimation to first estimate the velocities of MBs detected in a frame. The calculated velocities of subsequent frames are then used to estimate `tracks' within the frame; an example of the method is depicted in Fig. \ref{Fig:Prost_example}. As MBs flow inside blood vessels, they are more likely to be found in certain positions in the imaging plane. These tracks are then weighted in, per frame, and increase the likelihood for signal in their region to be detected as MBs. 

\subsection{Current limitations of super-resolution CEUS}
We now discuss some challenges of super-resolution US.
Complete imaging of vascular maps, and capillaries more so, requires that they be populated in some frames during scanning, yielding trade-offs between several desired properties: For high probability of population, it is beneficial to take a large number of frames. However, for conventional imaging schemes, a number of frames in the order of thousands, which is typically needed to populate organs such as the liver or the heart, requires several minutes of acquisition, creating discomfort, logistic complications and huge amounts of data, as well as increasing the probability of patient and organ motion mid-scan. Thus, a higher frame-rate is desired. This may be achieved by using plane-wave or diverging-wave imaging, which are faster and allow insonification of the entire scan region using a single pulse, allowing for ultrafast imaging with frame rates of up to hundreds of frames per second. However, these schemes suffer from poor resolution, causing an increased blurring of MBs and difficulty in localization.

The inherent 3D motion of organs and 3D structure of blood vessels also places a challenge and opportunity for 3D super-resolution imaging schemes to be developed. Currently, 3D imaging modes typically rely on sequential acquisition of 2D slices of a given volume, effectively forcing a projection of the 3D image on the imaging plane, one at a time, as is shown in \ref{fig:comm volumetric}. Methods relying on a complete data-set acquired using designated, 2D or 3D arrays, may be able to exploit the complex structure in the data in similar manners to those proposed in this section, while taking into account further complexities introduced by the higher dimensionality. One such complexity is the increase in data sizes, which will challenge the device's ability to acquire at reasonable frame-rate.

\section{Conclusions and Outstanding Challenges}
\label{sec:Conc}

In this review, we detailed common approaches for ultrasound imaging, and how their limitations can be overcome by modern signal processing and learning methods. In particular, we focused on super-resolution, reduction in channel-data rates and sparse-array design. For each, a mathematical model was described, as well as how it may exploit prior knowledge of the data structure to improve results. For example, using Fourier-domain knowledge of the signal and the pulse strucutre yields a reduction in sampling rate and thus data size. Modelling contrast agent sparsity in several domains, leads to super-resolved images in conditions previously not enabled such as high concentrations and low frame rates. Further advances in signal processing techniques, as well as the increasing popularity of unfolding algorithms using deep neural networks, may lead to leaps in performance in upcoming years, and also help address issues such as technician dependence and improved diagnostics.

There are, however, still several major challenges and opportunities, all of which can benefit from a signal processing point of view. 

Super-resolution imaging faces challenges in the fields of tissue suppression, organ motion detection and microbubble localization, with image and RF signal processing techniques constantly being developed and tested for different organs and scenarios. A development towards the direction of 3D super-resolution imaging may be necessary in order to overcome several limitations, such as the 3D nature of microvascular flow and organ motion. Moreover, complex imaging schemes designed to increase frame rate e.g. by compounding several, rapidly transmitted plane waves, are continuously developed and tested, paving the way to early, noninvasive diagnosis of inflammatory and cancerous pathologies.

3D scanning capability is a key topic of research in the US community even regardless of super-resolution. True, real-time, 3D imaging is considered one of the ultimate goals in ultrasound imaging and is challenging due to the shear amount of data needed to image typical volumes of interest such as a heart or bladder (see Fig. \ref{fig:comm volumetric}), and the large number of RF chains needed to convert the received pulses to digital. Combining sub-Nyquist techniques and 2D or 3D sparse arrays, may reduce the amount of data to feasible dimensions, enabling sufficiently high frame rates at sufficient imaging quality. This would in turn reduce the dependence on the operator, as the current, limited performance of 3D scanning requires great expertise in order to yield valuable images. Data size reduction will pave the way towards wireless, cloud-based medicine, where experts from around the world may provide real-time analysis to data gathered remotely by technicians, in real time, using low-cost, portable systems, as shown in \ref{fig:UScloud}.
Deep learning methods, developed directly on channel data, to target specific clinical applications and extraction of further clinical insights from channel data directly will allow unprecedented inference, regardless of data visualization constraints.

With advances in hardware, structure-based processing techniques, and the advent of deep learning
methods, ultrasound has the potential to offer imaging on the edge to patients worldwide at an affordable
price.

\bibliographystyle{IEEEtran}
\bibliography{general}

\end{document}